\documentclass[prd,%
preprint,%
tightenlines,showpacs,%
nofootinbib,%
amsfonts,amsmath]{revtex4}

\begin{document}

\title{Generalized Galileons:\\
All scalar models whose curved background extensions\\
maintain second-order field equations and stress tensors}

\author{C.~Deffayet} \email{deffayet@iap.fr}
\affiliation{AstroParticule \& Cosmologie,
UMR 7164-CNRS, Universit\'e Denis Diderot-Paris 7,
CEA, Observatoire de Paris,
10 rue Alice Domon et L\'eonie
Duquet, F-75205 Paris Cedex 13, France}

\author{S.~Deser} \email{deser@brandeis.edu}
\affiliation{Physics Department, Brandeis University, Waltham, Massachusetts
02454, USA, and Lauritsen Laboratory, California Institute of
Technology, Pasadena, California 91125, USA}

\author{G.~\surname{Esposito-Far\`ese}} \email{gef@iap.fr}
\affiliation{${\mathcal{G}}{\mathbb{R}}
\varepsilon{\mathbb{C}}{\mathcal{O}}$, Institut d'Astrophysique
de Paris, UMR 7095-CNRS, Universit\'e Pierre et Marie
Curie-Paris 6, 98bis boulevard Arago, F-75014 Paris, France}

\begin{abstract}
We extend to curved backgrounds all flat-space scalar field
models that obey purely second-order equations, while maintaining
their second-order dependence on both field and metric. This
extension simultaneously restores to second order the, originally
higher derivative, stress tensors as well. The process is
transparent and uniform for all dimensions.
\end{abstract}

\date{June 10, 2009}


\maketitle

\section{Introduction}
Recently, an interesting scalar field, ``Galileon'', theory
\cite{Nicolis:2008in}, inspired by the decoupling limit of the
Dvali-Gabadadze-Porrati (DGP) model \cite{Dvali:2000hr} and its
cosmological consequences~\cite{COSMODGP}, was
introduced.\footnote{Aspects of its phenomenology are studied in
Refs.~\cite{Nicolis:2008in,Chow:2009fm,Babichev:2009ee}.}
(This model was previously proposed in \cite{Fairlie}, also
in flat space, with a quite different motivation.)
Originally formulated in flat spacetime and dimension $D = 4$,
its defining property was that, while the action contains both
first and second derivatives, the equations of motion
\textit{uniquely} involve the latter. As shown in Ref.~\cite{US},
the simplest covariantization led to field equations for the
scalar $\pi$ and its stress tensor that contained third
derivatives; fortunately, \cite{US} also showed how to eliminate
these higher derivatives by introducing suitable nonminimal,
curvature, couplings. (This cure's small price was to break an
original symmetry of the model, that of shifting the first
derivatives of $\pi$ by a constant vector, which is not
meaningful in curved space anyhow.) Although the phenomenological
relevance of the nonminimal terms has not been studied,
\cite{US} furnished a nontrivial example of ``safe'', purely
second-order, class of scalar-tensor couplings. However, it was
restricted to $D=4$ and involved rather complicated algebra.

In the present work, we will provide the transparent and uniform
basis in arbitrary $D$ for this, \textit{a priori} surprising, nonminimal
completion. To do so, the Galileon model will first be
reformulated in Sec.~II; in particular, we will exhibit its
simplest flat-spacetime properties. Section III will incorporate
curved backgrounds, in $D=4$ for concreteness. This will
illustrate how the new formulation leads very directly to the
original nonminimal couplings of \cite{US}. The final section
completes our results by extending them to arbitrary dimensions
and backgrounds. Our results are encapsulated in Eqs.~(\ref{US1})
for flat, and (\ref{ActionCurvedBackground}) for general,
background.

To define our framework more precisely, we will exhibit, starting
from a transparent ``canonical'' flat-space action with purely
second-derivative field equation (but still unavoidably higher
derivative stress tensor), a ``minimal'' nonminimal
gravitational coupling extension that simultaneously guarantees
no higher than second derivatives of either field \textit{or}
metric in both the field equation \textit{and} stress tensor in
any $D$ and background. We do not claim uniqueness for this
construction simply because one may add infinitely many (rather
trivial because irrelevant) terms, all vanishing in flat space,
that also avoid higher derivatives. Examples include Lagrangians
such as (any function of) the scalar field times all
Gauss-Bonnet-Lovelock or Pontryagin densities, let alone plain
scalar curvature. Likewise, starting from a flat
``noncanonical'' version differing from ours by a total
divergence, other nonminimal terms would be generated. Finally,
our aim being to avoid higher than second derivatives, we will
not discuss, for us trivial, incidental first and zeroth order
terms such as $V(\pi)$.

\section{Flat-spacetime Galileon}
\label{SS1}
In Ref.~\cite{Nicolis:2008in} it was argued that the most general
flat-space action in $D$ dimensions for a scalar field $\pi$
whose field equations contain {\it only} second-order (but
neither zeroth, first, nor higher) derivatives is obtained by a
linear combination of the following Lagrangian
densities\footnote{In our notation, ${\cal L}_{(n,p)}$ is a
Lagrangian density that is a sum of monomials, each containing
products of $n$ fields $\pi$, acted on by first and second
derivatives, and $p$ explicit occurrences of the Riemann tensor.
Note that Eq.~(135) or (A4) of Ref.~\cite{Nicolis:2008in} equals $n$
times our Eq.~(\ref{nico1}).}
\begin{equation} \label{nico1}
{\cal L}_{(n+1,0)}= \sum_{\sigma \in S_n} \epsilon(\sigma)
\bigl[ \pi_{\vphantom{\mu_1}}^{\mu_{\sigma(1)}} \pi_{\mu_1}\bigr]
\bigl[\pi_{\hphantom{\mu_{\sigma(2)}} \mu_2}^{\mu_{\sigma(2)}}
\pi_{\hphantom{\mu_{\sigma(3)}} \mu_3}^{\mu_{\sigma(3)}} \ldots
\pi_{\hphantom{\mu_{\sigma(n)}} \mu_n}^{\mu_{\sigma(n)}}\bigr],
\end{equation}
where indices on the scalar field will always denote (ordinary or
covariant according to context) derivatives, e.g. $\pi_{\mu\nu}
\equiv \pi_{,\mu\nu}$ or $\pi_{;\mu\nu}$, and $\sigma$ denotes a
permutation of signature $\epsilon(\sigma)$ of the permutation
group $S_n$, with $n \leq D$. If this last inequality is not
satisfied, the above Lagrangian density (\ref{nico1}) vanishes
identically. Thus in four dimensions, there are only four
nontrivial Galileon Lagrangians (\ref{nico1}) beyond the
nonderivative ${\cal L}_{(1,0)}=\pi$ of
Ref.~\cite{Nicolis:2008in}; they are ${\cal L}_{(2,0)}=\pi_\mu
\pi^\mu$, a cubic Lagrangian ${\cal L}_{(3,0)}=\pi_\mu
\pi^\mu\Box \pi - \pi_\mu \pi^{\mu\nu}\pi_\nu = \frac{3}{2}
\pi_\mu \pi^\mu\Box \pi+ \text{tot. div.}$ (the one obtained in
the decoupling limit of DGP \cite{Dvali:2000hr}), and ${\cal
L}_{(4,0)}$ and ${\cal L}_{(5,0)}$:
\begin{eqnarray}
{\cal L}_{(4,0)} &=& \left(\Box \pi\right)^2
\left(\pi_{\mu}\pi^{\mu}\right)
-2 \left(\Box \pi\right)\left(\pi_{\mu}
\pi^{\mu\nu}\pi_{\nu}\right)
- \left(\pi_{\mu\nu}\pi^{\mu\nu}\right)
\left(\pi_{\rho}\pi^{\rho}\right)
+2 \left(\pi_{\mu}\pi^{\mu\nu}
\pi_{\nu\rho}\pi^{\rho}\right), \label{L4}\\
{\cal L}_{(5,0)} &=& \left(\Box \pi\right)^3
\left(\pi_{\mu}\pi^{\mu}\right)
-3 \left(\Box \pi\right)^2\left(\pi_{\mu}
\pi^{\mu\nu}\pi_{\nu}\right)
-3 \left(\Box \pi\right) \left(\pi_{\mu\nu}
\pi^{\mu\nu}\right) \left(\pi_{\rho}\pi^{\rho}\right)
\nonumber \\
&& +6 \left(\Box \pi\right)\left(\pi_{\mu}\pi^{\mu\nu}
\pi_{\nu\rho}\pi^{\rho}\right)
+2 \left(\pi_{\mu}^{\hphantom{\mu}\nu}
\pi_{\nu}^{\hphantom{\nu}\rho}
\pi_{\rho}^{\hphantom{\rho}\mu}\right)
\left(\pi_{\lambda}\pi^{\lambda}\right)
\nonumber \\
&& +3 \left(\pi_{\mu\nu}\pi^{\mu\nu}\right)
\left(\pi_{\rho}\pi^{\rho\lambda}\pi_{\lambda}\right)
-6 \left(\pi_{\mu}\pi^{\mu\nu}\pi_{\nu\rho}
\pi^{\rho\lambda}\pi_{\lambda}\right). \label{L5}
\end{eqnarray}
The Lagrangian (\ref{nico1}) can also be rewritten as
\begin{equation}
{\cal L}_{(n+1,0)}=\sum_{\sigma \in S_n} \epsilon(\sigma)
g^{\mu_{\sigma(1)}\nu_{\vphantom{()}1}}
g^{\mu_{\sigma(2)}\nu_{\vphantom{()}2}} \ldots
g^{\mu_{\sigma(n)}\nu_{\vphantom{()}n}}
(\pi_{\nu_1} \pi_{\mu_1})
(\pi_{\nu_2 \mu_2} \pi_{\nu_3 \mu_3}\ldots \pi_{\nu_n \mu_n}).
\end{equation}
As we will see, the key to success will be to rewrite the above
Lagrangians in terms of the totally antisymmetric Levi-Civita
tensor. We first recall the identity
\begin{equation}
\sum_{\sigma \in S_D} \epsilon(\sigma)
g^{\mu_{\sigma(1)}\nu_{\vphantom{()}1}}
g^{\mu_{\sigma(2)}\nu_{\vphantom{()}2}} \ldots
g^{\mu_{\sigma(D)}\nu_{\vphantom{()}D}} =
- \varepsilon^{\mu_{\vphantom{()}1} \mu_{\vphantom{()}2}
\ldots \mu_{\vphantom{()}D}}\,
\varepsilon^{\nu_{\vphantom{()}1} \nu_{\vphantom{()}2}
\ldots \nu_{\vphantom{()}D}},
\end{equation}
valid for any space and dimension, using
\begin{equation}
\varepsilon^{\mu_{\vphantom{()}1} \mu_{\vphantom{()}2} \ldots
\mu_{\vphantom{()}D}} = - \frac{1}{\sqrt{-g}}
\delta^{[\mu_{\vphantom{()}1}}_1 \delta^{\mu_{\vphantom{()}2}}_2
\ldots \delta^{\mu_{\vphantom{()}D}]}_D,
\end{equation}
where the square bracket denotes unnormalized permutations. From
our two $\varepsilon$ tensors, it is useful to define the
$2n$-contravariant tensor $\mathcal{A}_{(2n)}$ by contracting
$D-n$ indices:
\begin{equation} \label{DEFE}
\mathcal{A}_{(2n)}^{\mu_{\vphantom{()}1} \mu_{\vphantom{()}2}
\ldots \mu_{\vphantom{()}2n}} \equiv
\frac{1}{(D-n)!}\,
\varepsilon^{\mu_{\vphantom{()}1}
\mu_{\vphantom{()}3} \mu_{\vphantom{()}5} \ldots
\mu_{\vphantom{()}2n-1}\, \nu_{\vphantom{()}1}
\nu_{\vphantom{()}2}\ldots
\nu_{\vphantom{()}D-n}}_{\vphantom{\nu_{\vphantom{()}1}}}
\,\varepsilon^{\mu_{\vphantom{()}2} \mu_{\vphantom{()}4}
\mu_{\vphantom{()}6} \ldots
\mu_{\vphantom{()}2n}}_{\hphantom{\mu_{\vphantom{()}2}
\mu_{\vphantom{()}4} \mu_{\vphantom{()}6} \ldots
\mu_{\vphantom{()}2n}}\nu_{\vphantom{()}1}
\nu_{\vphantom{()}2}\ldots \nu_{\vphantom{()}D-n}}.
\end{equation}
The numerical factor $1/(D-n)!$ is introduced so that
$\mathcal{A}_{(2n)}$ keeps the same expression in terms of
products of metric tensors in any dimension $D\geq n$. To further
simplify notation, we sometimes replace indices $\mu_i$ by their
index $i$ whenever $i<10$ (but never larger, reinstating $\mu_i$
if needed). For example, (\ref{DEFE}) now reads
\begin{equation}
\mathcal{A}_{(2n)}^{1234\ldots } =
\frac{1}{(D-n)!}\,
\varepsilon^{135\ldots
\,\nu_{\vphantom{()}1} \nu_{\vphantom{()}2}\ldots
\nu_{\vphantom{()}D-n}}
\,\varepsilon^{246\ldots}_{\hphantom{246\ldots}
\nu_{\vphantom{()}1} \nu_{\vphantom{()}2}\ldots
\nu_{\vphantom{()}D-n}}.
\end{equation}
Note that the tensor $\mathcal{A}_{(2n)}$ is obviously
antisymmetric upon permutations of the odd ($1,3,5, \ldots$), as
well as those of even ($2,4,6, \ldots$), indices. Also, we will
only write expressions containing $\mathcal{A}_{(2n)}$ with all
indices up, and we will then omit those indices with the
convention that lower indices denoted by integers $1,2, \ldots,9$
or by indices $\mu_{i}$ are always contracted with the
corresponding upper ones of $\mathcal{A}_{(2n)}$. Hence, we will
use a letter different from $\mu$ to denote indices that are not
contracted with those of $\mathcal{A}_{(2n)}$. It is now easy to
see that the Lagrangian (\ref{nico1}) can be rewritten as
\begin{equation} \label{US1}
{\cal L}_{(n+1,0)}=-\mathcal{A}_{(2n)} (\pi_1 \pi_2)
(\pi_{34} \pi_{56} \pi_{78} \ldots
\pi_{\mu_{2n-1} \mu_{2n}}),
\end{equation}
while for example Lagrangians ${\cal L}_{(4,0)}$ and ${\cal
L}_{(5,0)}$ given in Eqs. (\ref{L4}) and (\ref{L5}) can be
rewritten in the compact form
\begin{eqnarray}
{\cal L}_{(4,0)} &=& -\varepsilon^{\mu_1 \mu_3 \mu_5 \nu}
\,\varepsilon^{\mu_2 \mu_4 \mu_6}_{\hphantom{\mu_1
\mu_3 \mu_5}\nu}\,
\pi_{\mu_1} \pi_{\mu_2}\, \pi_{\mu_3 \mu_4} \pi_{\mu_5 \mu_6}
= -\mathcal{A}_{(6)}\, \pi_{1} \pi_{2}\,\pi_{34} \pi_{56}, \\
\label{S51}
{\cal L}_{(5,0)} &=& -\varepsilon^{\mu_1 \mu_3
\mu_5 \mu_7} \,\varepsilon^{\mu_2 \mu_4 \mu_6 \mu_8}\,
\pi_{\mu_1}\pi_{\mu_2}\,\pi_{\mu_3 \mu_4} \pi_{\mu_5 \mu_6}
\pi_{\mu_7 \mu_8}
= -\mathcal{A}_{(8)}\, \pi_{1}\pi_{2}\,\pi_{34} \pi_{56}
\pi_{78}. \label{S52}
\end{eqnarray}
Clearly, the field equations derived from (\ref{US1}) only
contain second derivatives. Indeed, first, upon varying the
Lagrangian (\ref{US1}) with respect to $\pi$, the twice-differentiated
term appearing there gives rise, after integration
by parts, to third and fourth order derivatives acting on $\pi$,
of the form $\pi_{\mu_i\mu_j\mu_k}$ and
$\pi_{\mu_i\mu_j\mu_k\mu_l}$. But any expression of the form
$\mathcal{A}_{(2n)} \pi_{\mu_i\mu_j\mu_k}$ or $\mathcal{A}_{(2n)}
\pi_{\mu_i\mu_j\mu_k\mu_l}$ vanishes identically, because
flat-spacetime derivatives commute and such an expression
contains at least two indices among $\{i, j, k\}$ having the same
parity and hence contracted with the same epsilon tensor arising
in the definition of $\mathcal{A}_{(2n)}$. So not only does the
Lagrangian (\ref{US1}) lead to equations with at most second
derivatives, but it also means that when a term with a
twice-differentiated $\pi$ is varied, one must distribute, after
integrating by parts, its two derivatives onto the $\pi_{1}$ and
$\pi_{2}$ terms. Similarly, when either single derivative factor
is varied, that derivative must land only on the other, yielding
the only contribution, $\pi_{12}$. Hence, as announced, the field
equations arising from the variation of (\ref{US1}) contain only
second derivatives. They read $(n+1){\cal E}_{(n+1,0)}=0$, where
\begin{equation}
{\cal E}_{(n+1,0)} =-\sum_{\sigma \in S_n} \epsilon(\sigma)
\prod_{i=1}^{i=n} \pi_{\hphantom{\mu_{\sigma(i)}}
\mu_i}^{\mu_{\sigma(i)}}
= \mathcal{A}_{(2n)}
\pi_{12} \pi_{34} \pi_{56} \ldots \pi_{\mu_{2n-1} \mu_{2n}}.
\end{equation}

\section{Galileons in $D=4$ curved space}
\label{SS2}
In Ref. \cite{US}, it was noted that minimal covariantization of
(\ref{nico1}), just with covariant derivatives (still omitting
semicolons),
\begin{equation}
-\int d^D x \sqrt{-g}\, \mathcal{A}_{(2n)} (\pi_1 \pi_2)
(\pi_{34}\pi_{56} \pi_{78} \ldots
\pi_{\mu_{2n-1} \mu_{2n}}),
\end{equation}
led to third derivatives of the metric, as gradients of
curvatures, in the field equation, as well as to third
derivatives of $\pi$ in the stress tensor. This is not very
desirable, due to the well-known stability problems caused by
higher derivatives in both scalar and gravitational sectors: More
initial conditions would have to be specified, and in some
backgrounds, new excitations might appear. Note that these
problems arise as soon as the Lagrangians (\ref{nico1}) contain a
product of at least two twice-differentiated $\pi$'s, as will be
seen in detail in Sec.~\ref{SS3}. For example, in $D=4$, this is
the case for $\{{\cal L}_{(4,0)}, {\cal L}_{(5,0)}\}$, but not
for $\{{\cal L}_{(2,0)}, {\cal L}_{(3,0)}\}$. A way out was
provided in \cite{US} where it was shown that, in $D=4$, there
exists a unique (in the ``minimal'' sense explained in the
Introduction) nonminimal term that removes all the third
derivatives arising in both variations of the action: the field
equations and the stress tensor. Indeed, adding the Lagrangians
${\cal L}_{(4,1)}$ and ${\cal L}_{(5,1)}$,
\begin{eqnarray} {\cal L}_{(4,1)} &=&
\left(\pi_{\lambda}\pi^{\lambda}\right)
\pi_{\mu}\Bigl[R^{\mu \nu} - \frac{1}{2} g^{\mu \nu} R\Bigr]
\pi_{\nu},\\
{\cal L}_{(5,1)}&=& -3
\left(\pi_{\lambda}\pi^{\lambda}\right) \left(\pi_{\mu}
\pi_{\nu} \pi_{\rho\sigma} R^{\mu\rho\nu\sigma}\right)
-18 \left(\pi_{\mu}\pi^{\mu}\right)
\left(\pi_{\nu}\pi^{\nu\rho}R_{\rho\sigma}\pi^{\sigma}
\right) \nonumber \\
&& +3 \left(\pi_{\mu}\pi^{\mu}\right) \left(\Box \pi\right)
\left(\pi_{\nu} R^{\nu \rho}\pi_{\rho}\right) +
\frac{15}{2} \left(\pi_{\mu}\pi^{\mu}\right)
\left(\pi_{\nu} \pi^{\nu\rho}\pi_{\rho}\right) R
+\text{tot. div.}, \label{L51}
\end{eqnarray}
respectively to ${\cal L}_{(4,0)}$ and ${\cal L}_{(5,0)}$, we
obtain covariant Galileon actions whose field equations contain
derivatives of order lower or equal to two, both in $\pi$ and
metric variations. We now show how the nonminimal terms ${\cal
L}_{(4,1)}$ and ${\cal L}_{(5,1)}$ can easily be obtained using
our generalized form (\ref{US1}). To match the expressions for
${\cal L}_{(n+1,1)}$ derived below, a total derivative must
actually be added to Eq.~(\ref{L51}), namely 3 times Eq.~(18) of
Ref.~\cite{US}, which reads
\begin{eqnarray}
\text{tot. div.}&=& 3\left(\pi_\mu\pi^\mu\right)
\left(\pi_\nu\pi^\nu\right) \left(\pi_{\rho\sigma}
R^{\rho\sigma}\right)
+ 12 \left(\pi_\mu\pi^\mu\right)
\left(\pi_\nu\pi^{\nu\rho}
R_{\rho\sigma}\pi^\sigma\right)\nonumber\\
&&-\frac{3}{2} \left(\pi_\mu\pi^\mu\right)
\left(\pi_\nu\pi^\nu\right) \left(\Box \pi\right) R
-6 \left(\pi_\mu\pi^\mu\right)
\left(\pi_\nu \pi^{\nu\rho}\pi_\rho\right) R.
\label{totdiv51}
\end{eqnarray}

Let us first consider ${\cal L}_{(5,0)}$, and vary its action
with respect to $\pi$. Denoting it by $\delta_\pi {\cal
L}_{(5,0)}$, we have
\begin{equation} \label{DL53}
\delta_\pi {\cal L}_{(5,0)} = -2 \mathcal{A}_{(8)} \delta \pi_{1}
\pi_{2} \pi_{34} \pi_{56}\pi_{78} - 3 \mathcal{A}_{(8)}
\pi_{1} \pi_{2} \delta \pi_{34} \pi_{56}\pi_{78},
\end{equation}
where the coefficients 2 and 3 are easily obtained by a
renumbering of the dummy indices $\mu_i$. Upon integration by
parts, we see that the first term in the right-hand side above
cannot possibly lead to derivatives in the field equations of
order higher than two, because such terms could only (after
integration by parts) lead to third-order covariant derivatives
acting on $\pi$. But we know by construction that third
derivatives are absent in flat spacetime; hence they can only
lead, in curved backgrounds, to terms proportional to
(undifferentiated) curvatures times a first derivative of $\pi$.
The highest order derivatives appearing in such a product are
obviously of second order and act on the metric. Hence, the only
term which can potentially lead in the equations of motion to
derivatives of order higher than 2 (we will call those terms
``dangerous'' in the following) is
\begin{equation} \label{DL53zero}
\delta_\pi {\cal L}_{(5,0)} \sim -3 \mathcal{A}_{(8)} \pi_{1}
\pi_{2} \delta \pi_{34} \pi_{56}\pi_{78},
\end{equation}
where a tilde will mean that we only write the dangerous terms
and omit the others. Note that no dangerous terms are generated
by varying the volume factor $\sqrt{-g}$ in the action, so we may
henceforth work at Lagrangian density level and allow integration
by parts when writing expressions containing the $\sim$ symbol,
with the understanding that such expressions might differ by a
total derivative. When integrating the term on the right-hand
side of the above equation (\ref{DL53zero}) by parts to obtain
the $\pi$ field equation, we see, for reasons similar to those
given above, that the only dangerous terms occur when letting the
two derivatives, $\nabla_{\mu_3}$ and $\nabla_{\mu_4}$, act on an
already twice-differentiated $\pi$. We obtain
\begin{equation}
\label{DL53bis}
\delta_\pi {\cal L}_{(5,0)} \sim -3 \times 2 \,\delta \pi
\mathcal{A}_{(8)} \pi_{1} \pi_{2} \pi_{5643}
\pi_{78},
\end{equation}
where the extra factor 2 comes from the possibility that those
derivatives act on $\pi_{56}$ or $\pi_{78}$, both of which give
the same term, after appropriate renumbering and index
permutations. Using similar rearrangements, we can rewrite
(\ref{DL53bis}) as
\begin{eqnarray}
\label{DL53ter}
\delta_\pi {\cal L}_{(5,0)} &\sim&
-3 \,\delta \pi \mathcal{A}_{(8)}
\pi_{1} \pi_{2}\left( \pi_{5643}-
\pi_{5463}\right)\pi_{78} \label{step1} \nonumber \\
&\sim & -3 \,\delta \pi \mathcal{A}_{(8)} \pi_{1}
\pi_{2}\pi^{\lambda}R_{465\lambda;3} \pi_{78} \nonumber\\
&\sim& -\frac{3}{2} \,\delta \pi \mathcal{A}_{(8)} \pi_{1}
\pi_{2}\pi^{\lambda}\left(R_{465\lambda;3} +
R_{46\lambda3;5}\right)\pi_{78} \nonumber\\
&\sim&\frac{3}{2} \,\delta \pi \mathcal{A}_{(8)} \pi_{1}
\pi_{2}\pi^{\lambda}R_{3546;\lambda}\pi_{78}, \label{step4}
\end{eqnarray}
where the last line uses the Bianchi identity
$R_{46[35;\lambda]}=0$. Hence, as already shown in \cite{US}, the
$\pi$ field equations contain third derivatives of the metric, as
first derivative of the curvature. The above term is the only
dangerous one coming from the variation of
\begin{equation}
\label{SS53}
\int d^4x \sqrt{-g}\, {\cal L}_{(5,0)}.
\end{equation}
It can be cancelled by adding to the above action the following
\begin{equation}
\frac{3}{4} \int d^4x \sqrt{-g}\, \mathcal{A}_{(8)} \pi_{1}
\pi_{2}\left(\pi_{\lambda} \pi^{\lambda}\right) R_{3546}\,
\pi_{78},
\end{equation}
which on the other hand is easily seen not to generate any
further dangerous term. In fact one can check explicitly that
this action is identical to the one obtained from ${\cal
L}_{(5,1)}$, that is
\begin{equation}
\label{SS51}
{\cal L}_{(5,1)} = \frac{3}{4} \mathcal{A}_{(8)} \pi_{1}
\pi_{2}\left(\pi_{\lambda} \pi^{\lambda}\right) R_{3546}\,
\pi_{78}.
\end{equation}
It was shown in \cite{US} that the metric variation of
the sum (\ref{SS53}) plus (\ref{SS51}) does not contain
derivatives of order higher than two, but as we will see in the
next section, this can also easily be checked explicitly using
our expressions ${\cal L}_{(5,0)}$ and ${\cal L}_{(5,1)}$. Before
proceeding, let us note that a calculation similar to the one
given above leads to a simple expression for the nonminimal term
\begin{equation}
{\cal L}_{(4,1)} = \frac{1}{4} \mathcal{A}_{(6)}
\pi_{1} \pi_{2}\left(\pi_{\lambda} \pi^{\lambda}\right)
R_{3546}.
\end{equation}

\section{Arbitrary $D$ backgrounds}
\label{SS3}
We now show how the previous results can be generalized from
$D=4$ to arbitrary $D$. Namely, we will show that a covariant
Galileon model whose field equations have derivatives of order
lower or equal to two can be obtained in arbitrary dimensions by
a suitable linear combination of Lagrangians densities of the
type
\begin{equation} \label{Lnp}
{\cal L}_{(n+1,p)} = -\mathcal{A}_{(2n)}
\pi_{1} \pi_{2} \mathcal{R}_{(p)}
\mathcal{S}_{(q)},
\end{equation}
where $\mathcal{R}_{(p)}$ and $\mathcal{S}_{(q)}$ are defined by
\begin{eqnarray}
\mathcal{R}_{(p)} &\equiv& \left(\pi_{\lambda} \pi^{\lambda}\right)^p\,
\prod_{i=1}^{i=p}
R_{\mu_{4i-1} \;\mu_{4i+1}\;\mu_{4i}\;\mu_{4i+2}},\\
\mathcal{S}_{(q)} &\equiv& \prod_{i=0}^{i=q-1} \pi_{\mu_{2n-1-2i}\;
\mu_{2n-2i}},
\end{eqnarray}
and one has $q= n-1-2p$. The Lagrangian densities ${\cal
L}_{(n+1,p)}$ are obtained from ${\cal L}_{(n+1,0)}$ by replacing
$p$ times a pair of twice-differentiated $\pi$, by a product of
Riemann tensors by $\pi_{\lambda}\pi^{\lambda}$ (with suitable
indices). To further streamline the discussion and the notations,
we will also use (in the spirit of Petrov notation) an index
$A_i$ to denote the four indices $\mu_{4i-1} \,\mu_{4i+1}\,
\mu_{4i}\, \mu_{4i+2}$ taken in that order: We will write, e.g.,
\begin{equation}
\mathcal{R}_{(p)} = \left(\pi_{\lambda} \pi^{\lambda}\right)^p\,
\prod_{i=1}^{i=p} R_{A_i},
\end{equation}
and we will also use the convention that $\mathcal{R}_{(p)}$ and
$\mathcal{S}_{(q)}$ vanish respectively for $p < 0$ and $q < 0$,
while $\mathcal{R}_{(0)} = \mathcal{S}_{(0)} \equiv 1$ by
consistency of definition (\ref{Lnp}) with Eq.~(\ref{US1}). Let
us first look at the variation of ${\cal L}_{(n+1,p)}$, denoted
by $\delta_\pi {\cal L}_{(n+1,p)}$, with respect to $\pi$. We
find
\begin{eqnarray}
\delta_\pi {\cal L}_{(n+1,p)} &=& -2 \mathcal{A}_{(2n)} \delta
\pi_{1} \pi_{2} \mathcal{R}_{(p)} \mathcal{S}_{(q)} \nonumber\\
&&- 2 p \mathcal{A}_{(2n)} \pi_{1}\pi_{2} \mathcal{R}_{(p-1)}
\delta\pi_{\lambda}\pi^{\lambda}R_{A_p}
\mathcal{S}_{(q)} \nonumber\\
&&- q \mathcal{A}_{(2n)} \pi_{1}\pi_{2} \mathcal{R}_{(p)}
\delta \pi_{\mu_{4p+3}\; \mu_{4p+4}} \mathcal{S}_{(q-1)}.
\label{dLTOT}
\end{eqnarray}
After integrating by parts, the first term on the right-hand side
of the above equation does not lead to dangerous terms (in the
terminology of the previous section). Indeed, the only possible
dangerous terms it could generate are derivatives of the
curvature in the form $R_{A_i;1}$. However, when contracted with
$\mathcal{A}_{(2n)}$ those terms vanish by virtue of the Bianchi
identity $R_{\mu\nu [\rho \sigma;\kappa]} =0$. The terms obtained
from the second one of Eq.~(\ref{dLTOT}) by letting (after
integration by parts) the derivative $\nabla_\lambda$ act on
$\mathcal{S}_{(q)}$ are \textit{a priori} dangerous, because the index
$\lambda$ is not contracted with one index of
$\mathcal{A}_{(2n)}$ and hence our previous argument for
discarding third derivatives would fail. However, those terms are
exactly compensated (up to nondangerous ones) by those obtained
from an integration by parts of the third term of
Eq.~(\ref{dLTOT}), where the derivatives $\nabla_{\mu_{4p+3}}
\nabla_{\mu_{4p+4}}$ act on one of the $\pi_\lambda$ of
$\mathcal{R}_{(p)}$. We thus find, by a rewriting similar to
(\ref{step4}), that the dangerous terms in the variation
$\delta_\pi {\cal L}_{(n+1,p)}$ read
\begin{eqnarray}
\delta_\pi {\cal L}_{(n+1,p)} &\sim& 2 p^2 \mathcal{A}_{(2n)}
\pi_{1} \pi_{2} \mathcal{R}_{(p-1)} \pi^{\lambda}R_{A_p;\lambda}
\mathcal{S}_{(q)} \nonumber \\
&&+\frac{q(q-1)}{4} \mathcal{A}_{(2n)} \pi_{1} \pi_{2}
\mathcal{R}_{(p)} \pi^{\lambda}R_{A_{p+1};\lambda}
\mathcal{S}_{(q-2)}. \label{DP}
\end{eqnarray}
Note that this expression also holds for $p=0$ and $q=0$, $q=1$.

Let us now consider the variation $\delta_g {\cal L}_{(n+1,p)}$
of ${\cal L}_{(n+1,p)}$ with respect to the metric. Defining the
variation of the metric $g_{\mu \nu}$ by $h_{\mu \nu}$, those of
$\pi_{\mu_{4p+3} \, \mu_{4p+4}}$ and of $R_{A_p}$, denoted by
$\delta_g \pi_{\mu_{4p+3} \, \mu_{4p+4}}$ and $\delta_g R_{A_p}$,
respectively obey
\begin{eqnarray}
\delta_g \pi_{\mu_{4p+3} \, \mu_{4p+4}} &=& -\frac{1}{2}
\pi^{\sigma}\left( h_{\sigma \,\mu_{4p+4}\,;\,\mu_{4p+3}} +
h_{\sigma\, \mu_{4p+3}\,;\,\mu_{4p+4}}-
h_{\mu_{4p+3}\,\mu_{4p+4}\,;\,\sigma}\right),\\
\mathcal{A}_{(2n)} \delta_g R_{A_p} &=& 2 \mathcal{A}_{(2n)}
h_{\mu_{4p-1}\,\mu_{4p+2}\,;\,\mu_{4p+1}\,\mu_{4p}}
+ \mathcal{A}_{(2n)}h^\sigma_{\mu_{4p-1}}
R_{\sigma\,\mu_{4p+1}\,\mu_{4p}\,\mu_{4p+2}}.
\end{eqnarray}
{}From those equations, it follows that $\delta_g {\cal
L}_{(n+1,p)}$ contains the dangerous terms
\begin{eqnarray}
\delta_g {\cal L}_{(n+1,p)} \sim &&
\frac{q(q-1)}{2} \mathcal{A}_{(2n)} \pi_{1} \pi_{2}
\mathcal{R}_{(p)} \pi^{\sigma}\pi_{\mu_{4p+5}\, \mu_{4p+6}\,
\sigma}\, \mathcal{S}_{(q-2)} h_{\mu_{4p+3}\,
\mu_{4p+4}} \nonumber\\
&& + \frac{p q}{2}
\mathcal{A}_{(2n)} \pi_{1} \pi_{2} \mathcal{R}_{(p-1)}
\pi^{\sigma} R_{A_p;\sigma} \pi_{\lambda}\pi^{\lambda}
\mathcal{S}_{(q-1)} h_{\mu_{4p+3}\,
\mu_{4p+4}} \nonumber\\
&& - 2 p q \mathcal{A}_{(2n)} \pi_{1} \pi_{2}
\mathcal{R}_{(p-1)} \pi_{\lambda}\pi^{\lambda}
\pi_{\mu_{4p+3}\,\mu_{4p+4}\,\mu_{4p}\,\mu_{4p+1}}
\mathcal{S}_{(q-1)} h_{\mu_{4p-1}\,\mu_{4p+2}}\nonumber\\
&& - 4 p^2 \mathcal{A}_{(2n)} \pi_{1} \pi_{2}
\mathcal{R}_{(p-1)} \pi^{\lambda}
\pi_{\lambda\,\mu_{4p}\,\mu_{4p+1}}\,
\mathcal{S}_{(q)} h_{\mu_{4p-1}\,\mu_{4p+2}}.
\end{eqnarray}
{}From a rewriting again similar to (\ref{step4}), it is easily
seen that the second and third terms on the right-hand side of
the above equation cancel each other. Then, after some relabeling
and permutation of dummy indices, one is left with
\begin{eqnarray}
\delta_g {\cal L}_{(n+1,p)} \sim && \frac{q(q-1)}{2}
\mathcal{A}_{(2n)} \pi_{1} \pi_{2} \mathcal{R}_{(p)}
\pi^{\lambda}\pi_{\mu_{4p+5}\, \mu_{4p+6}\, \lambda}\,
\mathcal{S}_{(q-2)} h_{\mu_{4p+3}\, \mu_{4p+4}} \nonumber\\
&& +4 p^2 \mathcal{A}_{(2n)} \pi_{1} \pi_{2}
\mathcal{R}_{(p-1)} \pi^{\lambda}
\pi_{\lambda\,\mu_{4p+2}\, \mu_{4p+1}}
\mathcal{S}_{(q)} h_{\mu_{4p-1}\,\mu_{4p}}. \label{Dg}
\end{eqnarray}
Using the above expressions (\ref{DP}) and (\ref{Dg}), it is then
easy to see that the action given by
\begin{equation}\label{ActionCurvedBackground}
I = \int d^D x \sqrt{-g} \sum_{p=0}^{p_{\rm max}}
\mathcal{C}_{(n+1,p)} {\cal L}_{(n+1,p)},
\end{equation}
with $p_{max}$ the integer part of $(n-1)/2$ [i.e., the number of
pairs of twice-differentiated $\pi$ in $\mathcal{S}_{(n-1)}$],
leads to field equations (both for $\pi$ and the stress tensor)
with no more than second derivatives, provided the coefficients
$\mathcal{C}_{(n+1,p)}$ satisfy the recurrence relation
\begin{equation}
\mathcal{C}_{(n+1,p)} = - \frac{(n+1-2p)(n-2p)}{8\,p^2}
\,\mathcal{C}_{(n+1,p-1)}.
\end{equation}
The latter is easily solved by (setting $\mathcal{C}_{(n+1,0)}$
to one)
\begin{equation}
\mathcal{C}_{(n+1,p)} = \left(-\frac{1}{8}\right)^p
\frac{(n-1)!}{(n-1-2p)!\,(p!)^2}
= \left(-\frac{1}{8}\right)^p
\binom{n-1}{2p}\binom{2p}{p}.
\end{equation}
These coefficients correspond to those of $(xy)^p$ in the
expansion of $(1+x-y/8)^{n-1}$. Remarkably, they suffice to
ensure the disappearance of dangerous terms in both the metric
and $\pi$ field equations.

\section{Conclusions}
We have presented, in arbitrary $D$ and gravitational
backgrounds, the ``minimally'' most general scalar models whose
field equations and stress tensors depend on second field
derivatives and (undifferentiated) curvatures. Whatever their
ultimate physical usefulness, it is remarkable that these models
exist at all and even more that they can be systematized in so
uniformly simple a manner. Their construction is tantalizingly
reminiscent of gravitational Gauss-Bonnet-Lovelock models.

\section*{Acknowledgments}
We thank the authors of Ref. \cite{Nicolis:2008in} and J.~Mourad
for discussions, and D.~Fairlie for bringing Refs.~\cite{Fairlie} to our
attention. The work of S.D. was supported by NSF Grant No.~PHY
07-57190 and DOE Grant No.~DE-FG02-92ER40701.

\end{document}